\documentstyle[12pt,psfig]{article}

\begin{document}
\baselineskip=24 true pt

\title{Spin-Reversal Transition in Ising Model under Pulsed Field}
\author{Arkajyoti Misra and Bikas K. Chakrabarti\\
Saha Institute of Nuclear Physics\\
1/AF Bidhannangar,\\
Calcutta 700 064,\\
India.}
\maketitle

\begin{abstract}
In this communication we report the existence of a dynamic ``spin-reversal''
transition in an Ising system perturbed by a pulsed external magnetic field. 
The transition is achieved by tuning the strength ($h_p$) and/or the duration 
($\Delta t$) of the pulse which is applied in a direction opposite to the 
existing order. We have studied this transition in the kinetic Ising Model
in two dimension using Monte Carlo technique, and solved numerically the mean 
field equation of motion.
The transition is essentially dynamic in nature and it takes
the system from one ordered equilibrium phase to another by means of the 
growth of opposite spin domains (in the kinetic Ising case) induced during 
the period when the pulsed field is applied.
\end{abstract}

\newpage

\section{Introduction}

\noindent
The dynamic response of Ising systems under positive pulsed fields
has recently been studied extensively employing computer simulation.
In particular, some interesting divergent growth of relaxation time and
finite time scaling behaviour was observed near the order-disorder transition
point of the Ising model for ``positive pulses'', where the pulsed field in the
ordered phase is in the direction of order \cite{abc}. Here we report the computer
simulation study of the same system where the pulsed field is ``negative'', i.e
in opposition to the existing order having equilibrium magnetization $m_0$,
below the order-disorder transition point. Due to the application
of a negative pulse $h_p$ in an ordered equilibrium phase,
the down spin domains start growing as long as the pulse is present.
Depending on the number and size of these down spin domains at the time
of withdrawal of the pulse, either they further grow to reach the other
equivalent ordered phase with reversed magnetization  ($-m_0$) or reduce to settle 
down in the original ordered phase with magnetization ($m_0$). The relaxation time 
$\tau$ taken by the system to reach either of these equilibrium states depends
on the strength ($h_p$) and  duration ($\Delta t$) of the pulse, as well 
as on the temperature $T$.

It may be mentioned that a number of studies have been made 
(see for example \cite{bind-prl},\cite{stauffer}) on the growth
of negative spin domains in a kinetic Ising model when an ordered state 
is suddenly subjected to a negative field, where the time variation of the 
field is step function like. The applicability of the classical
theory of nucleation by Becker and D\"{o}ring \cite{gunton} has also been
investigated extensively here. However, our problem differs from these
studies because the applied field is withdrawn after a finite interval of
time. The minimum amplitude of the field required to trigger a spin reversal
is non-vanishing if $\Delta t$ is finite. As we go on to increase the
pulse width, in the limit $\Delta t \rightarrow \infty$ we recover the
results for a field of `step-function' like nature (in time) where even an 
infinitesimally
small amplitude of the field is sufficient to trigger the spin-reversal.

  We have studied the ``phase diagram'' in the 
$h_p - \Delta t$ plane for the spin-reversal transition (from $m_0(T)$ to $-m_0(T)$) 
in a Monte Carlo simulation using Glauber Dynamics \cite{binder} and also from
the mean field equation of motion. In the kinetic Ising case we have also studied 
the variation or growth of the
relaxation time $\tau$ as one approaches this phase boundary. We observe clear
divergence of the relaxation time as the phase boundary is approached, 
indicating the spin-reversal transition to be a clear thermodynamic transition
with a divergent correlation length.

\noindent
\section{ Model and Simulation}

\noindent
In the Monte Carlo study, we take an Ising system of size $200 \times 200$
on a square lattice with nearest neighbour interaction without any disorder.
The Hamiltonian of the system is 

\begin{equation}
H = -J \sum_{\langle i j \rangle} S_i S_j - h(t) \sum_i S_i ~~,
\end{equation}

\noindent
where $S_i=\pm 1$ represents the Ising spins at lattice site $i$ and $J$ 
denotes the nearest neighbour interaction strength. The time dependent external
magnetic field $h(t)$ is applied as a pulse for a short duration $\Delta t$ :

\begin{eqnarray}
h(t) & = - h_p , & {\rm for} ~~~t_0 < t < t_0+\Delta t \nonumber \\
     & = 0 , &  {\rm otherwise} .
\end{eqnarray}

\noindent
The pulsed field is applied after the original Ising system comes to an 
equilibrium ordered phase (corresponding to $T < T_c$, the order-disorder
transition temperature). The time $t_0$ in (2) is thus much larger 
than the relaxation time of the pure Ising system without any perturbation. 
The direction of the 
pulsed field ($-h_p$) is opposite to the existing order or average 
equilibrium magnetization $m_0(T)$ existing at times before $t_0$.

Before the pulsed magnetic field is applied, the 
system is brought to equilibrium which is characterized by temperature only.
The system evolves according to the Glauber single spin flip dynamics. 
One complete sweep through the entire lattice is defined as one
Monte Carlo step (MCS) or one unit of time $t$. After the system 
reaches its equilibrium the 
pulse of strength $h_p$ is applied and is withdrawn after a finite time
interval $\Delta t$. After that the system is left to itself to
come to equilibrium and is attracted or evolves towards either of the two 
equally likely equilibria (determined by the temperature $T$) having different 
order parameter values or magnetization. Unless the system
is above the critical temperature of the unperturbed system
($T_c \simeq 2.27$), majority of the spins will point to a particular direction.
Suppose, before
the application of the pulsed field the magnetization of the system is 
$+m_0(T)$.  Now either by tuning the pulse width $\Delta t$ or the pulse 
height $h_p$ we can perturb the system 
in such a manner that after the withdrawal of the pulse the system chooses
to go over to the other equilibrium state, characterized by the magnetization 
$-m_0(T)$. We call it a ``spin-reversal'' transition when the sign of 
equilibrium magnetization is flipped by the application of the negative 
pulse. We have studied in our Monte Carlo calculations the phase diagram for 
such spin-reversal transition in the $h_p-\Delta t$ plane at a fixed 
temperature ($T<T_c$).  We have also looked at the relaxation behaviour of the 
dynamics of such systems (in particular the relaxation time $\tau$ for the 
average magnetization) as one approaches the phase boundary.
Typical number of samples (Monte Carlo seeds) taken for averaging the data 
points is 10.

We have also solved numerically the mean field equation of motion 
\begin{equation}
\frac{dm(t)}{dt} = -m(t) + \tanh\left(\frac{m(t) + h(t)}{T}\right),
\end{equation}
where $m(t)$ is the magnetization (per site) at time $t$ and $h(t)$ 
is given by (2).

\section{Results}

\noindent
It is quite obvious that all possible combinations of $h_p$ and $\Delta t$ cannot give rise to 
the spin-reversal transition at a particular temperature. Fig. 1 shows how the 
transition can be brought about by increasing $\Delta t$ for a fixed value of $h_p$ at 
a constant temperature. Fig. 2 shows similar effect by increasing $h_p$ while keeping
$\Delta t$ and $T$ constant.

At any particular temperature ($T<T_c$) there exists a combination of $h_p$ and 
$\Delta t$ which just manages to induce the spin-reversal transition. This $h_p-\Delta t$ 
curve is given by the phase diagram shown in Fig. 3. The inner side or
the axes side of the curves corresponds to the original phase, whereas one gets
"spin-reversed" phase for all the combinations of $h_p$ and $\Delta t$ outside the
phase boundary. The limitation arising out of the discrete time
simulations force the value of $\Delta t$ to start from 1, i.e one MCS.
With this kind of technique for estimating the 
critical value of pulsed field strength $h_p(\Delta t)$, the estimate of the 
phase boundary for $\Delta t<1$ is not possible. The phase boundaries tend to 
touch the abscissa at large values of $\Delta t$. This is well anticipated 
because 
even with an infinitesimally small strength, a negative field will eventually 
give rise to spin-reversal if applied for sufficiently long time.

	An important observation can be made from the series of figures shown in
Fig. 4. In Figs. 4(a)-(c) the time series plots of $m(t)$ are shown when $h_p$
is increased (at fixed $\Delta t$ and $T$) to reach the phase boundary from below. 
In Figs. 4(d)-(f) the phase boundary is approached from above by decreasing $h_p$
at the same value of $\Delta t$ and $T$. In either case,
it is clear that as one approaches the phase boundary for a particular
temperature and pulse width, the (relaxation) time taken by the system to reach 
its final equilibrium state increases. 

This prompts us to define the relaxation time $\tau$ of the system
as the time (MCS) taken by it to reach the final equilibrium state from the time of 
withdrawal of the pulse. In Fig. 5, we look for the variation of the quantity 
$\tau$ with $h_p$ for a particular temperature and fixed pulse width. It is 
clearly seen that $\tau$ seems to diverge as we approach the phase boundary 
from either side for that particular temperature and pulse width. Similar 
growth of $\tau$ is also observed while approaching the phase boundary by
varying $\Delta t$ at fixed $T$ and $h_p$. Such 
divergent growths of relaxation time clearly indicate the thermodynamic nature 
of the spin-reversal transition (divergent correlation length). It may be 
mentioned here that due to very large scatter (sometimes by order of magnitude) 
in the values of $\tau$ for different Monte Carlo seeds (otherwise 
thermodynamically identical samples) 
, log averaging turned out to be a better choice than ordinary averaging of $\tau$. The data for $\tau$ shown in Fig. 5 are obtained using log-averages
for $\tau$, thereby keeping the relative error less than $O(10^{-2})$.

The mean-field phase diagram is shown in fig. 6. Since there is no fluctuation,
there exists a finite  coercive field. The spin reversal does not occur even
for infinite pulse width ($\Delta t \rightarrow \infty$), if the pulse height
$h_p$ does not exceed the coercive field. The coercive field $h_p(\infty)$
decreases as $(T_c - T)^{\frac{3}{2}}$ with increasing temperature in the mean 
field case, where $T_c = 1$ in (3). This is because $m_0 h_p(\infty) \sim $
free energy $ F(m_0) \sim (T-T_c)m_o^2 + O(m_0^4) \sim (T-T_c)^2$.
 The inset of Fig. 6 shows the variation of 
$h_p(\infty)$ as a function of temperature. Unlike the kinetic Ising case, the
mean field phase diagrams can be extended beyond $\Delta t$ = 1. However since
the $\tanh$ function saturates to negative unity (for large $h_p$), the spin reversal can not
occur by further increasing the magnitude of the negative field if it is not 
applied for sufficient time. From (3) we can write $dm = [-m + \tanh \beta(m - h_p)]dt
\approx -(m_0 + 1)dt$. Now $\int_0^{\Delta t} dm \approx -(m_0 + 1)\Delta t$ should
be sufficient to make the value of $m$ decrease from $m_0$ to 0. At low 
temperatures ($m_0 \simeq 1$) it requires $\Delta t \simeq 1$ while for higher
temperatures spin reversal occurs for $\Delta t {< \atop \sim} 1$.

\section{Discussions}
Using Monte Carlo simulations for Ising system evolving under Glauber
dynamics (with non-conserving order parameter), we have studied a new dynamic 
``spin-reversal'' transition, where the system goes from one stable equilibrium
to another due to the application of a pulsed field (of finite duration) opposite
to the existing order of the system. We have determined the phase diagram
for such a transition in the pulse strength ($h_p$)-pulse width ($\Delta t$) plane 
for a fixed temperature $T$ less than the order disorder transition temperature
$T_c$. We observe that the typical relaxation time $\tau$ tends to diverge
as the phase boundary is approached, indicating a divergent correlation 
length associated with such dynamic transition.

In the kinetic Ising Model, the transition is actually triggered by the 
eventual growth of the ``negative''
spin domains formed during the period when the negative field was ``on''. 
According to the classical nucleation theory, number of droplets of size
$l$ is given by
\[ n_l = N \exp(-\epsilon_l /T) \]
where $N$ is a normalization constant. Here the free energy
for formation of a droplet is given by
\[ \epsilon_l = 2h_pl + \sigma l^{\frac{d-1}{d}} \]
in $d$ dimensions, where $\sigma$ is proportional to the surface tension. From the 
optimality of $\epsilon_l$ the estimated nucleation rate from 
Becker-D\"{o}ring theory is given by
\begin{equation}
I = I_0 exp({-\epsilon_{l_c}/T});~~~
l_c = \left( \frac{\sigma (d - 1)}{2 d h_p} \right)^d
\end{equation}
where $I_0$ is some constant depending on temperature.
Equating the rate $I$ with the inverse pulse width $\Delta t$ one gets 
approximately $h_p \sim 1/(\ln \Delta t)$ in $d = 2$ for the phase boundary. 
However, 
it can be checked from Fig. 3, this is not the case even for high temperature
phase boundaries. This is because the spin-reversal does not necessarily
have to take place during the presence of the field, it may occur long after
the withdrawal of the pulse. In fact, in our case
$\epsilon_{l_c}$ is expected to have also a $\Delta t$ dependence .
This can be clearly seen from Fig. 1, where the relaxation rate, after the withdrawal 
of the field, is strongly dependent on the pulse width ($\Delta t$). 

From the linearized limit of the mean field equation (3), one gets
\begin{equation}
m(t) = \left( m_0 - \frac{h_p}{1-T} \right) e^{(1-T)(t-t_0)/T} + \frac{h_p}{1-T},
\end{equation}
for $t>t_0$ and very close to $t_0$, so that linearization of (3) is possible.
Since there is no fluctuation in the mean field limit, there cannot be any
spin reversal if the magnetization remains positive at the time of withdrawal
of the pulse. Demanding that the magnetization should at least be zero at the
time of withdrawal of the field for an eventual  spin reversal, we find the relation
between $h_p$ and $\Delta t$ at the phase boundary :
\begin{equation}
\Delta t = \left( \frac{T}{1-T}\right) \ln \left( \frac{h_p}{h_p+(T-1)m_0}\right).
\end{equation}
For temperatures close to $T_c=1$, equation (6) can be approximated as
\begin{equation}
h_p \Delta t \simeq m_0 T,
\end{equation}
which indeed compares fairly well with the mean field phase boundaries (cf. Fig. 6)
in the region long before saturation.

The detailed study of the nature of the domain growth in this case of pulsed 
fields and their statistics are in the process, and shall be published elsewhere.

\section*{Acknowledgement}

We would like to thank  A.K. Sen and  D. Chowdhury for some useful discussions.
We are also thankful to D. Stauffer for valuable comments and suggestions.

\newpage
\section*{Figure Captions}

{\bf Figure 1.} Time series plots of the pulse ($h(t)$) and the magnetization ($m(t)$)
for $T=1.0$ and $h_p=1.04$: (a) $\Delta t=10$ (b) $\Delta t=15$ (c) $\Delta t=28$.

\noindent
{\bf Figure 2.} Time series plots of the pulse ($h(t)$) and the magnetization ($m(t)$)
for $T=1.0$ and $\Delta t=10$: (a) $h_p=1.04$ (b) $h_p=1.17$ (c) $h_p=1.34$.

\noindent
{\bf Figure 3.} Phase diagram (Monte Carlo) in the $h_p-\Delta t$ plane for
$T$ = 0.1 ($\bullet$), 0.5($\nabla$), 1.0($\Box$), 1.5($\triangle$),
2.0($\ast$). The typical errors present in the data points are less than
the symbol sizes.

\noindent
{\bf Figure 4.} Time series plots of the pulse ($h(t)$) and the magnetization ($m(t)$) 
for $T=1.5$ and $\Delta t=40$: (a) $h_p=0.40$ (b) $h_p=0.45$ (c) $h_p=0.47$ 
(d) $h_p=0.55$ (e) $h_p=0.49$ (f) $h_p=0.48$.

\noindent
{\bf Figure 5.} The behaviour of $\tau$ as the phase boundary is approached from either
side: (a) $T=0.50,\Delta t=10$ (b) $T=1.00,\Delta t=2$ (c) $T=2.00,\Delta t=10$.

\noindent
{\bf Figure 6.} Phase diagram (mean field) in the $h_p-\Delta t$ plane
for $T$ = 0.1, 0.3, 0.5, 0.8, 0.9. Inset : Variation of $h_p(\infty)$ with T.

\newpage

\begin{figure}
\psfig{file=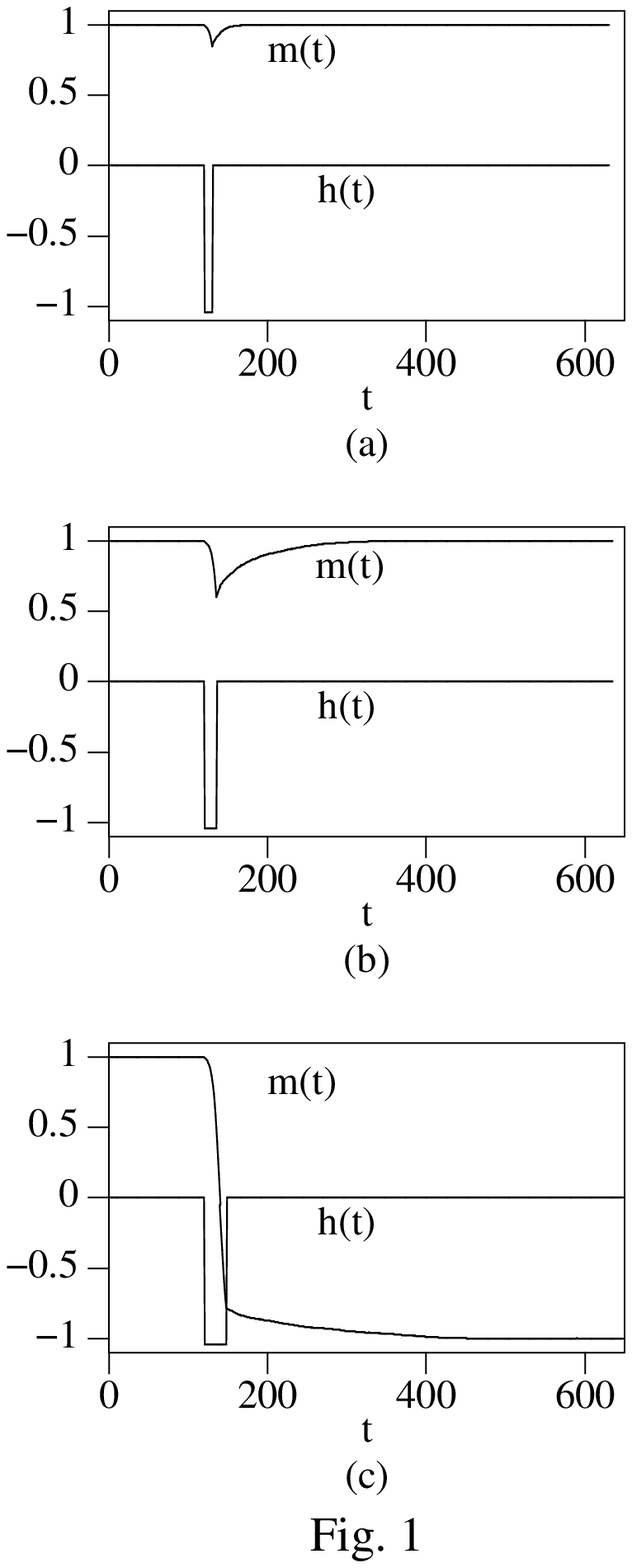} 
\end{figure} 

\clearpage

\begin{figure}
\psfig{file=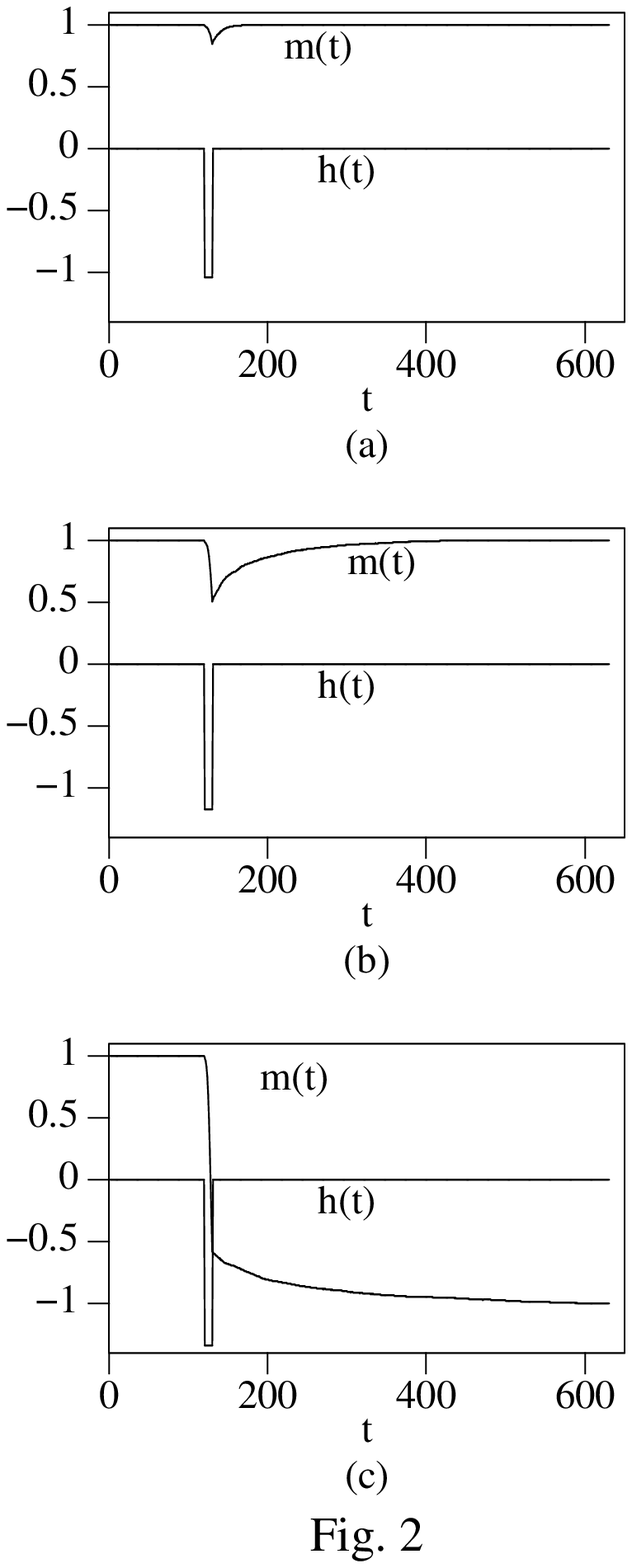}
\end{figure}

\clearpage

\begin{figure}
\psfig{file=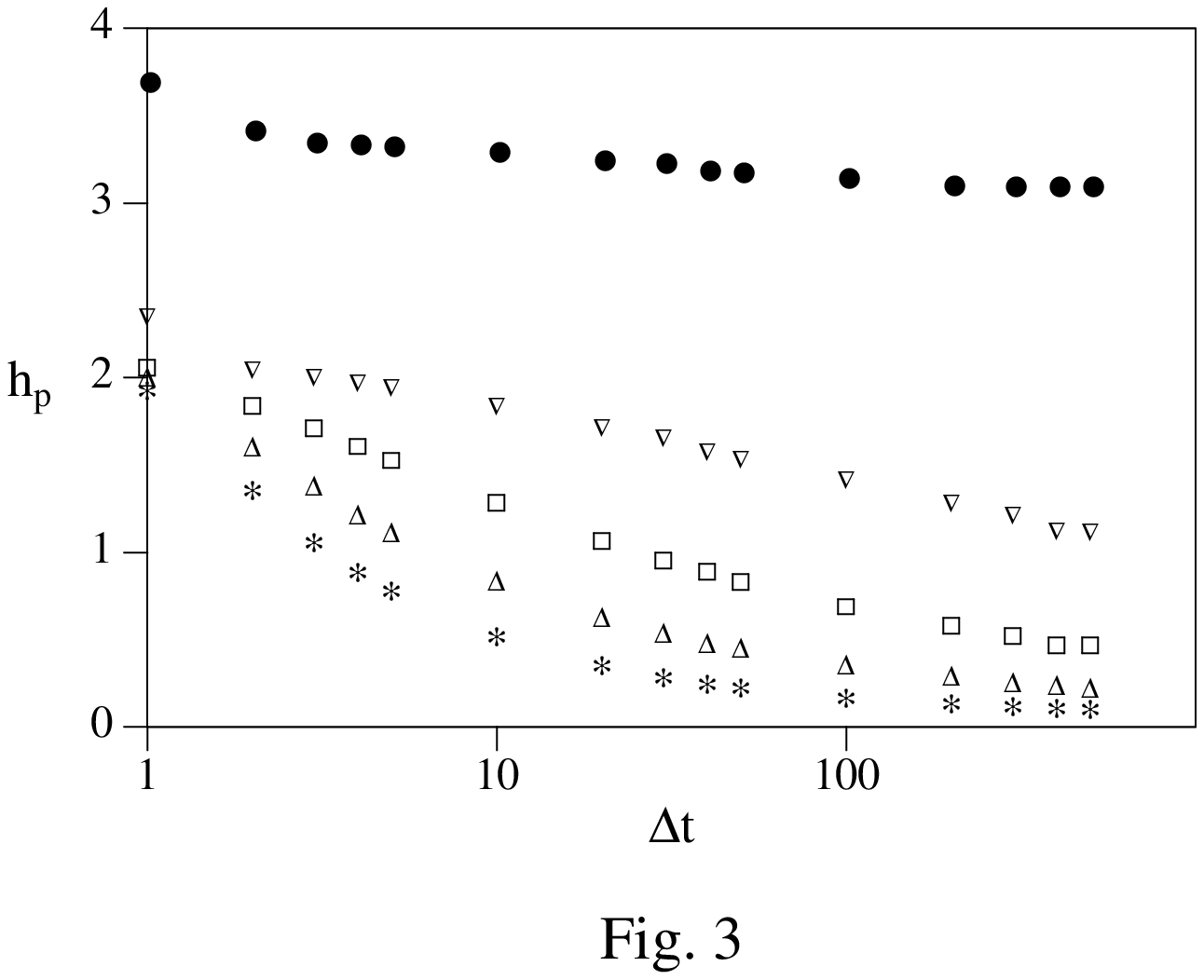}
\end{figure}

\clearpage

\begin{figure}
\psfig{file=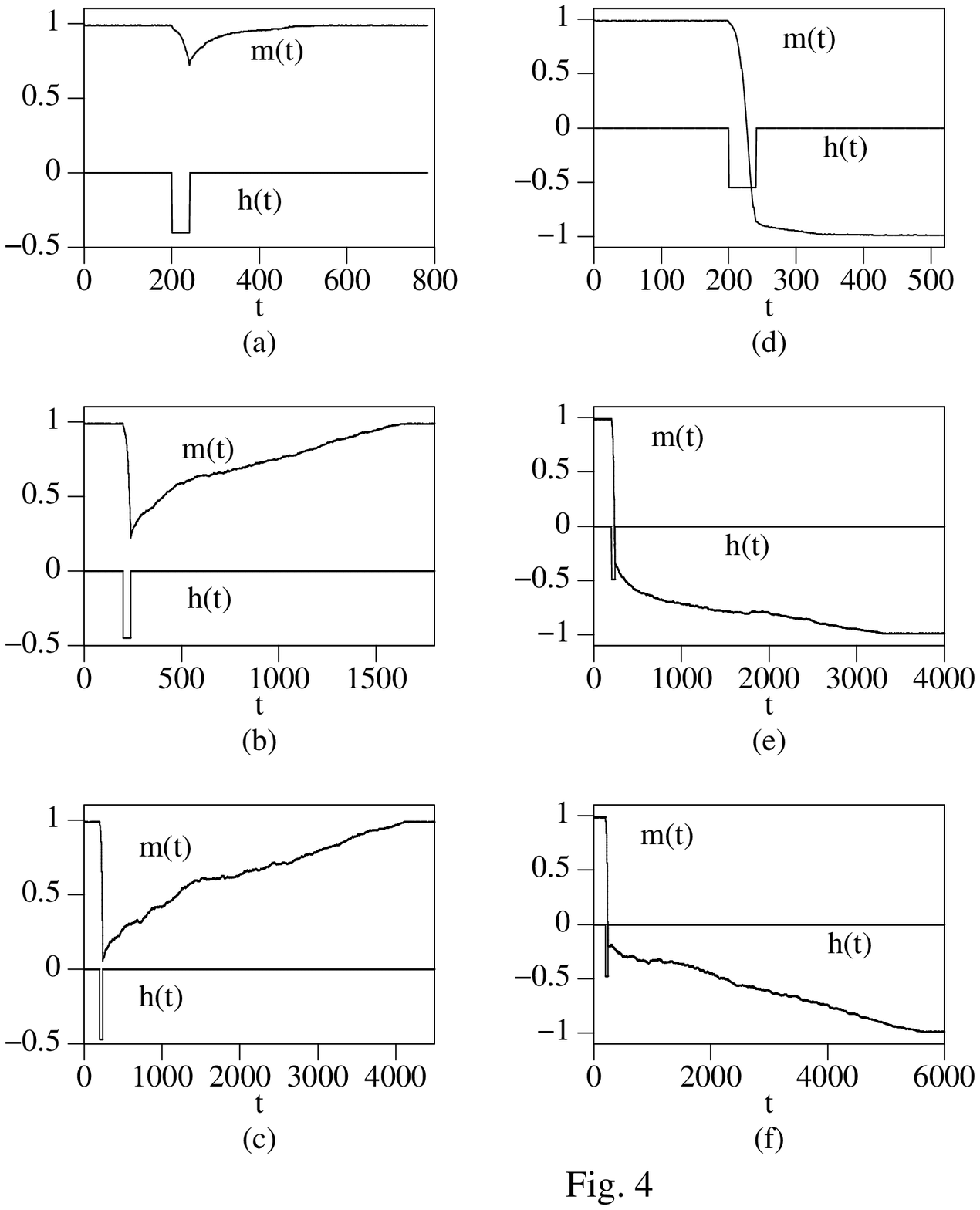}
\end{figure}

\clearpage

\begin{figure}
\psfig{file=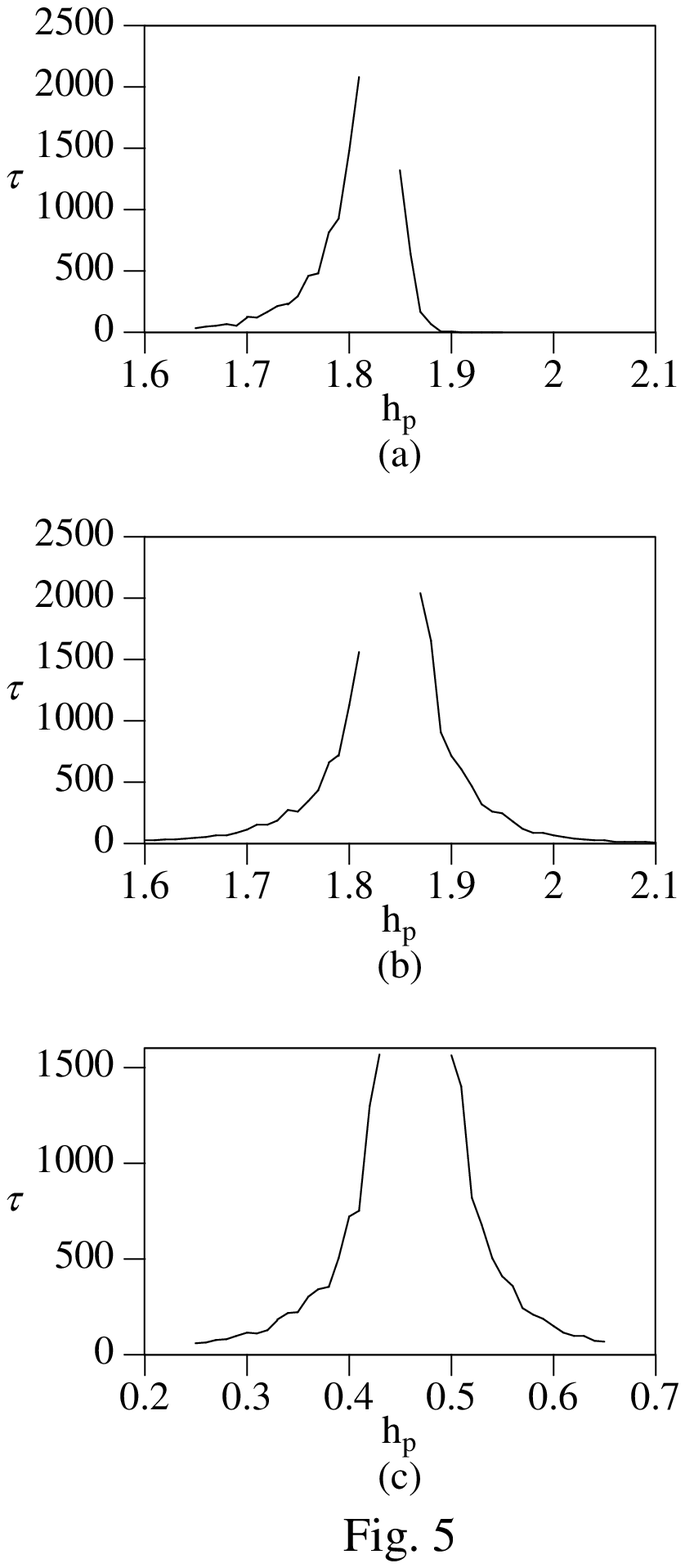}
\end{figure}

\clearpage

\begin{figure}
\psfig{file=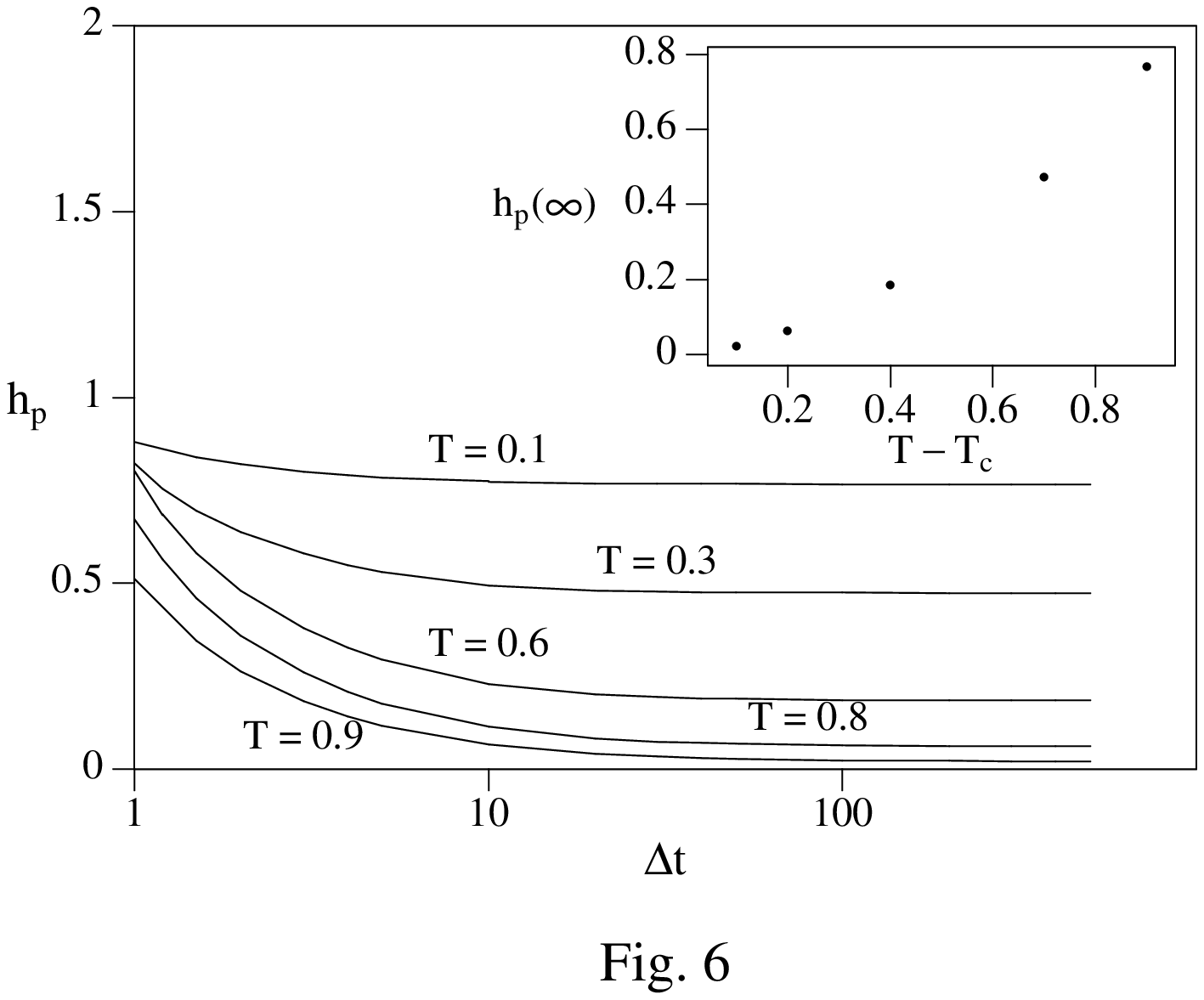}
\end{figure}


\newpage

\begin{thebibliography}{10}

\bibitem{abc} M. Acharyya, J. K. Bhattacharjee and B. K. Chakrabarti, Phys. Rev. E
{\bf 55} (1997) 2392

\bibitem{bind-prl} K. Binder and E. Stoll, Phys. Rev. Lett. , {\bf 31} 
(1973) 47

\bibitem{stauffer} D. Stauffer, Int. J. Mod. Phys. C,
{\bf 3} (1992) 1059

\bibitem{gunton} {\em Introduction to the Theory of Metastable and Unstable
States}, J. D. Gunton and M. Droz, Lecture Notes in Physics (183), Springer-Verlag,
Heidelberg (1983)

\bibitem{binder} {\em Application of the Monte Carlo method in Statistical Physics},
Ed. K. Binder, Springer, Heidelberg (1984)

\end{thebibliography}
\end{document}